# Dirac-Electrons-Mediated Magnetic Proximity Effect in Topological Insulator / Magnetic Insulator Heterostructures

Mingda Li[1,2*†], Qichen Song[1†], Weiwei Zhao[3], Joseph A Garlow[4], Te-Huan Liu[1], Lijun Wu[4], Yimei Zhu[4], Jagadeesh S. Moodera[5,6], Moses H. W. Chan[3], Gang Chen[1*] and Cui-Zu Chang[3,5*]

[1]*Department of Mechanical Engineering, MIT, Cambridge, MA 02139, USA*
[2]*Department of Nuclear Science and Engineering, MIT, Cambridge, MA 02139, USA*
[3]*Department of Physics, The Pennsylvania State University, University Park, PA16802, USA*
[4]*Condensed Matter Physics and Materials Science Department, Brookhaven National Lab, Upton, New York 11973, USA*
[5]*Francis Bitter Magnet Lab, MIT, Cambridge, MA 02139, USA*
[6]*Department of Physics, MIT, Cambridge, MA 02139, USA*

The possible realization of dissipationless chiral edge current in a topological insulator / magnetic insulator heterostructure is based on the condition that the magnetic proximity exchange coupling at the interface is dominated by the Dirac surface states of the topological insulator. Here we report a polarized neutron reflectometry observation of Dirac electrons mediated magnetic proximity effect in a bulk-insulating topological insulator $(Bi_{0.2}Sb_{0.8})_2Te_3$ / magnetic insulator EuS heterostructure. We are able to maximize the proximity induced magnetism by applying an electrical back gate to tune the Fermi level of topological insulator to be close to the charge neutral point. A phenomenological model based on diamagnetic screening is developed to explain the suppressed proximity induced magnetism at high carrier density. Our work paves the way to utilize the magnetic proximity effect at the topological insulator/magnetic insulator hetero-interface for low-power spintronic applications.

Magnetic proximity effect (MPE) in a topological insulator (TI) / magnetic insulator (MI) heterostructure induces magnetization to the TI's electronic states driven by means of magnetic exchange coupling from the MI layer [1-14]. As long as the MPE induced magnetic order is realized through the Dirac electrons, this effect is expected to have applications in low-energy consumption electronic and spintronic devices [1,2,15-17]. One prominent application is the quantum anomalous Hall effect (QAHE), where dissipationless chiral edge current can be harbored without external magnetic field [18-24]. Comparing with the other approach to magnetize the Dirac surface state (DSS) by doping the TI with transitional metal ion doping [20,21,25-29], the MPE approach has the advantage that it can result in a uniform magnetization over the entire TI layer without creating any impurity sites nor destroying the nontrivial band structure [2,8,14,30]. While the QAHE has been realized in Cr- and V- doped TI systems [18-20,22,23,27], no QAHE has been hitherto demonstrated in a MPE induced ferromagnetism system in spite of extensive studies. This brings up a few fundamental questions: is the MPE actually happening to the DSS of TI? If so, what is the nature of the MPE induced ferromagnetic order in DSS? For the first question, MPE in principle can happen to any electronic states of TI, including bulk states [31]. As to the second question, if, for instance, the MPE is realized through coupling with free carriers of the TI, the MPE will be enhanced by increasing the carrier density, hence hampering QAHE due to its carrier-free requirement [21]. Therefore an experimental study on the correlation between the carrier density and the magnitude of MPE in TI will show whether the MPE through a MI is a viable route to realize QAHE.

In this *Letter*, by equipping a polarized neutron reflectometer (PNR) with additional electrical transport capability, we are able to determine the effect of bottom gate-voltage ($V_g$) on proximity induced magnetism at the interface between TI $(Bi_{0.2}Sb_{0.8})_2Te_3$ and MI (EuS). Specifically we found maximum proximity induced magnetism is realized when the Fermi level is close to the charge neutral point.

The PNR experiments were carried out at beamline NG-D at the NIST Center for Neutron Research (NCNR), at fixed temperature $T$=5K. PNR is a powerful technique for measuring the real-space magnetic structure of thin films [32]. The experimental setup is illustrated in Fig. 1b, where the incident spin-polarized neutron beams are reflected by the MI/TI heterostructure (red and blue spheres), while the spin non-flip reflectivity signals from spin-up and spin-down components, $R_\uparrow$ and $R_\downarrow$, are collected alternatively. The PNR is integrated with a custom-built sample holder (Fig. 1b and Supplemental Materials I), which allows *in-situ* gate-dependent two-terminal longitudinal transport measurement while performing the PNR measurements. A $\mu_0 H = 0.7$T guide field is exerted at all times to prevent possible neutron depolarization. Fig. 1a is a schematic of the EuS / $(Bi_{0.2}Sb_{0.8})_2Te_3$ heterostructure. Both

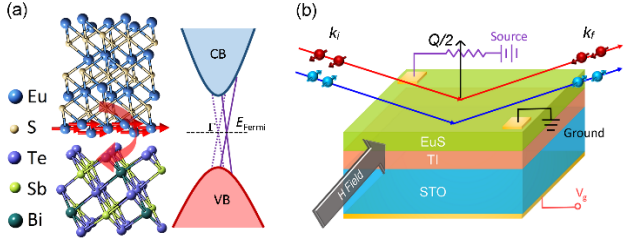

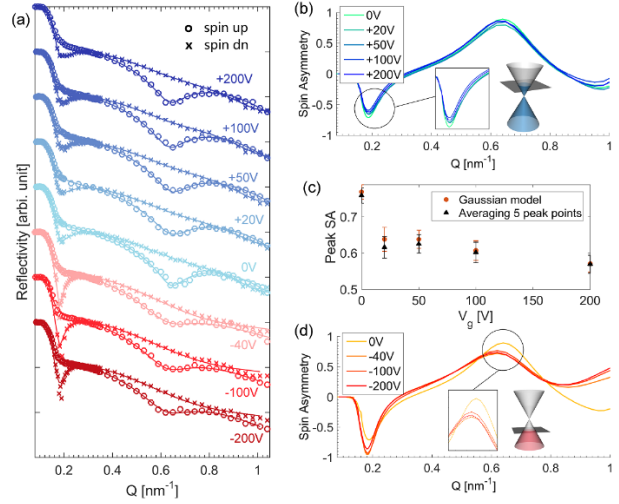

**Fig. 1** (color). (a) The atomic configuration (left) and schematic band structure (right) at the MI/TI interface. When the ferromagnetic Eu ions magnetize the surface state of the TI (red arrow in left figure), the original Dirac cone (purple dotted lines) is shifted away from Γ point (purple solid lines) through exchange coupling, causing MPE. (b) The schematics of the gate-dependent neutron reflectometry system, where the incident neutron momentum, reflected neutron momentum and the momentum change are denoted as $k_i$, $k_f$, and Q ( $Q \equiv k_i - k_f$), respectively. The two neutron spins switched by a spin flipper shine onto the sample alternatively. An image of this setup is shown in Supplemental Materials I.

perpendicular magnetization, which causes the surface state bandgap opening [21], and the in-plane magnetization which results in the shift of Dirac cone [24] are capable of inducing QAHE. In the present setup, we focus on the in-plane magnetization, yet there might still be canting effect which allows the spin rotation toward perpendicular direction [9]. The reason for choosing Eu-based element as proximity layer is its extremely high neutron absorption cross section ( $\sigma_A$(Eu) = 4530barn ), which helps to identify the location of the proximity layer and has been implemented in a few recent studies [8,9]. Moreover, an ambipolar behavior of MPE is revealed, that for both n- and p-doped TI, the MPE is reduced when the carrier density is increased. This strongly suggests that the MPE in a TI /MI heterostructure is not originated from the free carriers, hence enables the MPE based QAHE. The reduced proximity-induced magnetism at increased carrier density can be understood as a diamagnetic screening effect. We have developed a phenomenological model to explain this effect, qualitatively. Our study sheds light on further exploring the direction of MPE toward next generation dissipationless electronic and spintronic applications.

Since the purpose of this measurement is to detect the MPE's possible variation, which can be considered as a 2nd order effect if the ferromagnetism in MI layer is 0th order and the induced MPE is 1st order effect, the sample quality becomes essential. High-quality MI 5nm EuS / 4 quintuple layers (QLs) intrinsic TI $(Bi_{0.2}Sb_{0.8})_2Te_3$ heterostructure is grown by molecular beam epitaxy (MBE) under a base vacuum ~5×10$^{-10}$ Torr. The Bi:Sb ratio is carefully optimized to locate the Fermi level close to the Dirac point based on the method reported in [33]. The 4QL TI thickness is chosen make sure no hybridization gap is formed on the surface states and to facilitate the bias gate voltage tunability due to the low bulk carrier concentration [34]. The TI thin film is grown on top of heat-treated 0.5mm thick $SrTiO_3$ (STO) (111) substrate for back-gating purpose. The high sample quality is checked from both X-ray diffraction (Supplemental Materials I) and TEM (Fig. 3d), where high quality epitaxial growth is realized with sharp MI/TI interface.

**Fig. 2** (color). (a) The experimental measured (o and x dots) and theoretically refined (solid lines) PNR reflectivity (in log scale) vs. Q of both spin components $R_\uparrow$ and $R_\downarrow$, at various backgate voltages $V_g$=0V, +20V, +50V, +100V, +200V, -200V, -100V and -40V (based on the measuring sequence). Q is the magnitude of the neutron momentum change, where at low Q, $R_\uparrow = R_\downarrow = 1$. Each reflectivity curve is vertically shifted by $10^2$ for clarity. (b, d) The spin asymmetry (SA) at positive and negative $V_g$s, respectively. Even before any refinement, a higher spin asymmetry peak can be seen at low $V_g$, indicating a greater change of local magnetic structure at small $V_g$, as shown in (c).

The measured reflectivity curves at various $V_g$ are plotted in Fig. 2a, which are further refined by using the GenX program to extract the scattering length density (SLD) profiles [35]. The "Best-1-Bin" algorithm has been implemented during the differential evolution refinement process, indicating a likelihood to achieve global optimization within reasonable parameter range [36]. In fact, even prior to seeing the refined SLD, we could already tell the effect of voltage through raw data of spin asymmetry (SA), defined as SA $\equiv R_\uparrow - R_\downarrow / R_\uparrow + R_\downarrow$: at low $V_g$ near the charge neutral point (0V and +20V), the peak SA at 0V is higher and reduced at higher voltage, indicating a tunability of proximity through $V_g$ (Figs. 2b, 2c and Supplemental Materials II). The $V_g < 0$ part is not compared directly with the $V_g > 0$ part due to the sample heating and re-cooling between the measurements, to eliminate the STO memory effect when switching the $V_g$ polarity, but the gate-voltage dependent SA can still be seen clearly from (Fig. 2d and inset).

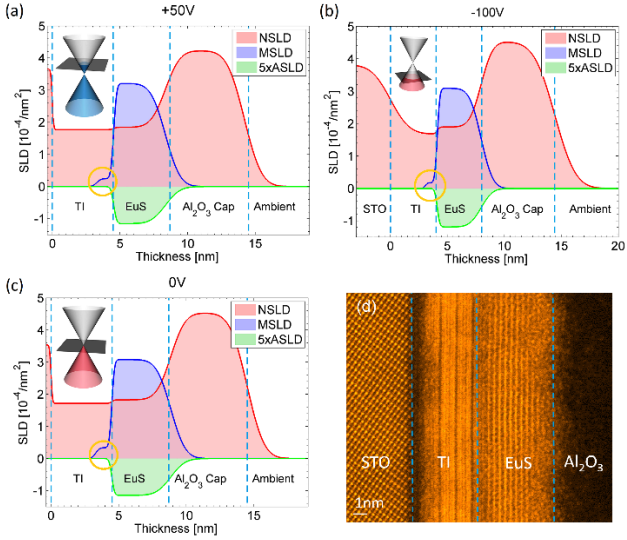

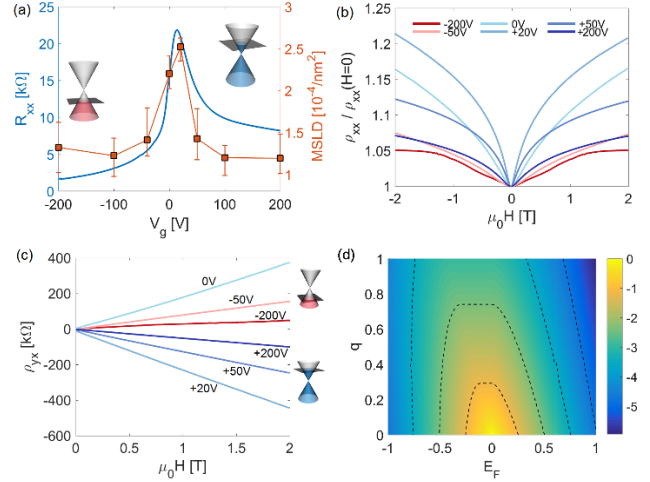

**Fig. 3** (color). (a-c) The SLD profiles at +50V, + 0V and -100V, respectively. The NSLD contains the information of sample thickness and interfacial roughness, which agrees well with the TEM result in (d). From the MSLD, we could see that the magnetic proximity effect (yellow circled region) can be tuned through $V_g$, and is reduced for both *n*- and *p*- type doping. The ALSD signal solely comes from the Eu ions, hence the lack of ASLD ~4nm in the TI side indicates that the proximity induced magnetism is not coming from the interdiffused Eu ions, which is further confirmed from the sharp TI/EuS interface in the TEM image. On the other hand, there are diffused Eu ions to the $Al_2O_3$ cap, which can be seen from both ASLD ~9nm at rough interface of $EuS/Al_2O_3$ from the TEM.

Upon fitting, the fitted curves (solid lines in Fig. 2a) show excellent agreement with the experiment data (filled points), with logarithmic figure of merit $<5\times10^{-2}$. To eliminate any hysteresis effect caused by the ferroelectricity of STO [37], the whole sample was warmed up to room temperature and cooled back down under high vacuum during the polarity change from +200V to -200V. For this reason, a very small thickness and density variation is allowed in the refinement process to obtain better fitting quality (Supplemental Materials III). A global fitting of all spectra simultaneously with strictly fixed thickness and SLD is performed independently using Refl1D package [38], showing an identical qualitative trend, that the proximity magnetism is higher at low gate voltage (Supplemental Materials IV).

The refined SLD profiles from the fitted reflectivity curves at a few representative $V_g$ are plotted in Fig. 3. The resulting nuclear scattering length density (NSLD) agrees well with the expected cross section calculation, that the STO substrate gives $\sim 3.6\times10^{-4} nm^{-2}$, TI layer $\sim 1.8\times10^{-4} nm^{-2}$, EuS $\sim 1.9\times10^{-4} nm^{-2}$ and amorphous $Al_2O_3$ capping layer $\sim 4\times10^{-4} nm^{-2}$. From the magnetic

**Fig. 4** (color). (a) The sheet longitudinal resistance $R_{xx}$ of the MI/TI heterostructure (blue curve) vs the proximity induced magnetization *M* (orange dots), as a function of gate voltage $V_g$. The proximity induced magnetism shows a peak near the charge neutral point, where the $R_{xx}$ also shows a peak. (b) Normalized longitudinal magnetoresistance $R_{xx}(H)$, from which a typical WAL behavior is seen in almost all bias voltages. (c) Hall resistance $R_{yx}(H)$ as a function of $V_g$. From the sign we could tell that at $V_g$ =0V it is hole-doped ($R_{yx}$>0), while at $V_g$ =20V it is electron-doped ($R_{yx}$<0). (d) The computed susceptibility $\chi$ as a function of Fermi level $E_F$ and wavenumber *q*. A diamagnetic effect $\chi<0$ is shown on both sides away from the Fermi level.

scattering length density (MSLD), it can be seen clearly that at $V_g$=+20V, the proximity MSLD is increased to $2.5\times10^{-4}/nm^2$, comparing with a $\sim 1.2\times10^{-4}/nm^2$ at $V_g$=+200V, which is comparable with previous report in [8], where TI $Sb_2Te_3$ was used. The warming and cooling of the sample between $V_g$>0 measurements and $V_g$<0 measurements causes some change of STO/TI interfacial roughness (Fig. 3 a vs b), making the fixed-SLD analysis valid for $V_g$>0 only (Supplemental Material III). Despite that two independent PNR refinement strategies and SA from raw data point to the same conclusion of peak proximity, we do not intend to fully exclude other possible SLD configurations due to the nuance effect caused by electric field.

At this stage, the effect of carrier density tuned through a $V_g$ to the proximity induced magnetism becomes clear. The MPE strength as a function of $V_g$ is plotted in Fig. 4a, together with the sheet longitudinal resistance $R_{xx}$. It can be seen clearly that the MPE shows a peak at $V_g$=0V, which coincides with the behavior of the $R_{xx}$, indicating that the highest MPE is realized near the charge neutral point. Moreover, a finite carrier density –both *n*-type or *p*-type – could reduce the MPE to the normal value as reported in [8]. On the other hand, the *ex-situ* magnetoresistance (MR) is performed as well. For longitudinal MR measurement (Fig. 4b), a typical weak anti-localization (WAL) behavior is revealed at almost all voltages [39,40]. In the weak $H$ regime, the MR shows a steeper increase as the Fermi level close to the charge neutral point. This behavior is consistent with the previous report [41]. In addition, the Hall measurements are also performed (Fig. 4c), from which we could verify the carrier type and estimate the carrier density [42]. The lowest carrier density $n_{2D}$= -2.8×10$^{12}$cm$^{-2}$ is located at $V_g$=+20V. Since the carrier densities at $V_g$=-200V, -50V, +50V and +200V are $n_{2D}$=2.64×10$^{13}$ cm$^{-2}$, 8.0×10$^{12}$ cm$^{-2}$, -5.0×10$^{12}$ cm$^{-2}$ and -2.75×10$^{13}$ cm$^{-2}$ respectively, while the maximum Dirac electron carrier density accommodated in the spin-polarized Dirac cone of a TI gives ~1.6×10$^{12}$cm$^{-2}$ (assuming $E_{gap}=0.3$eV and $v_F = 5\times 10^5$ m/s), we can see that the Fermi level is already tuned from the bulk valence band to bulk conduction band even at low $V_g$. This is consistent with the report in [43] that the critical carrier density of the Dirac cone is ~5.0×10$^{12}$ cm$^{-2}$, and is useful to simplify the theory discussed below.

To further understand the implication in Fig. 4a, we developed a simple phenomenological model to account for this ambipolar behavior from a perspective of magnetic susceptibility. It is well known that the bulk electronic bands of TI away from Dirac point have a diamagnetic nature due to large orbital diamagnetism [31], where the total magnetic susceptibility $\chi_{tot}$ under long-wavelength limit satisfies $\chi_{tot}(q \to 0) \sim -\frac{1}{3}\chi_s^B(q \to 0) < 0$, in which $\chi_s^B$ denotes the spin susceptibility for bulk bands. On the other hand, the DSS has a paramagnetic nature due to its small effective mass $\chi_{tot}(q \to 0) \sim +\frac{2}{3}\chi_s^D(q \to 0) > 0$, where $\chi_s^D$ denotes the spin susceptibility for DSS [31]. This indicates that whenever Fermi level is near the Dirac point (lower carrier density $n_{2D}$), the total susceptibility is enhanced through the paramagnetic effect of DSS. On the other hand, when the Fermi level is shifted away from the Dirac point (high $V_g$), the diamagnetism starts to dominate and screens the proximity magnetism. Assuming a conduction band minimum $E_c$ and valence band maximum $E_v$, the total susceptibility $\chi_{tot}$ as a function of Fermi level $E_F$ can thus be written as a phenomenological piecewise function

$$\chi_{tot}(E_F) \approx -\frac{1}{3}\chi_s^B(E_F)[\theta(E_F - E_c) + \theta(E_v - E_F)] \\ + \frac{2}{3}\chi_s^D(E_F)[\theta(E_F - E_v) - \theta(E_F - E_c)] \quad (1)$$

where $\theta(x)$ is the Heaviside step function. The spin-susceptibility $\chi_s$ can be written as [31]

$$\chi_s^{B,D}(\mathbf{q}) = \frac{\mu_B^2}{4\pi^3}\sum_{\substack{m,occ \\ n,empty}}\int d^3\mathbf{k}\frac{f_0(E_{n,\mathbf{k}}^{B,D}) - f_0(E_{m,\mathbf{k+q}}^{B,D})}{E_{m,\mathbf{k+q}}^{B,D} - E_{n,\mathbf{k}}^{B,D} + i\delta} \quad (2)$$

where $E_{n,\mathbf{k}}^{B,D}$ are the eigenvalues for the corresponding surface or bulk Hamiltonian, $f_0$ is the Fermi-Dirac distribution function. Here, by plugging the material-dependent eigenvalues [44] into Eq. (2), the susceptibility can be computed accordingly. For the particular EuS / (Bi$_{0.2}$Sb$_{0.8}$)$_2$Te$_3$ system, due to the lack of the corresponding effective Hamiltonian, here only a qualitative picture is provided, which is sufficient to demonstrate that whether carrier density can cause a unipolar ($\chi$ changes monotonically with $E_F$) or a ambipolar effect ($\chi(E_F = 0)$ is a local minimum or maximum). Since the bias voltage is shown to be effective enough to tune $E_F$ into bulk bands even at low $V_g$, assuming $E_v \approx E_c$ (small paramagnetic region) and neglecting all anisotropy which allows a much simpler dispersion relation $E_{m,\mathbf{k+q}} \sim E_{m,k+q} = \sqrt{k^2 + q^2}$ (resulted trend does not depend on the particular form of dispersion), a ambipolar effect is revealed (Fig. 4d), that when $E_F$ is away from charge neutral point, the diamagnetic screening gets stronger for both $n$-doping and $p$-doping, hence reduces the MPE. This agrees with our experimental results in Fig. 4a. At even higher $E_F$, the screening effect is saturated, since besides screening, other electronic states could directly contribute to the proximity-induced magnetism [31].

In summary, we showed the Dirac electrons mediated MPE at the interface of a MI/TI heterostructure by means of gate dependent PNR measurements. The tunability of the MPE enables future theoretical studies to utilize MPE to manipulate the interfacial spin texture in TI [45-47]. The maximum MPE, in particular, is realized when the Fermi level is approaching to the charge neutral point where carrier density is minimized. This proves that the MPE is favored at low carrier concentration, which opens up the possibility in realizing in-plane magnetization induced QAHE [24] at MI/TI heterostructure.

M.L., Q.S. and G.C. would like to thank support by S$^3$TEC, an Energy Frontier Research Center funded by U.S. Department of Energy (DOE), Office of Basic Energy Sciences (BES) under Award No. DE-SC0001299/DE-FG02-09ER46577, and DARPA MATRIX Program Contract HR0011-16-2-0041. M.L. and C.Z.C thank the helpful discussion of A. Grutter and B. Kirby. Y.Z. and L.W. acknowledge the support from DOE-BES / Materials Science and Engineering (MSE) division under Contract No. DE-AC02-98CH10886. JSM acknowledges the support from NSF Grants No. DMR-1207469, ONR Grant No. N00014-16-1-2657, and the STC Center for Integrated Quantum Materials under NSF Grant No. DMR-1231319.

C.Z.C would like to thank support from the startup grant provided by Penn State University.

†These authors contribute equally to this work.
* Authors to whom correspondence should be addressed:
mingda@mit.edu
gchen2@mit.edu
cxc955@psu.edu

# Supplemental Materials for
# Dirac-Electrons-Mediated Magnetic Proximity Effect in Topological Insulator / Magnetic Insulator Heterostructures

Mingda Li[1,2*†], Qichen Song[1†], Weiwei Zhao[3], Joseph A Garlow[4], Te-Huan Liu[1], Lijun Wu[4], Yimei Zhu[4], Jagadeesh S. Moodera[5,6], Moses. H. W. Chan[3], Gang Chen[1*] and Cui-Zu Chang[3,5*]

[1]Department of Mechanical Engineering, MIT, Cambridge, MA 02139, USA
[2]Department of Nuclear Science and Engineering, MIT, Cambridge, MA 02139, USA
[3]Department of Physics, The Pennsylvania State University, University Park, PA16802, USA
[4]Condensed Matter Physics and Materials Science Department, Brookhaven National Lab, Upton, New York 11973, USA
[5]Francis Bitter Magnet Lab, MIT, Cambridge, MA 02139, USA
[6]Department of Physics, MIT, Cambridge, MA 02139, USA

## I. Real image of gate-dependent measurement and XRD characterization of the heterostructure

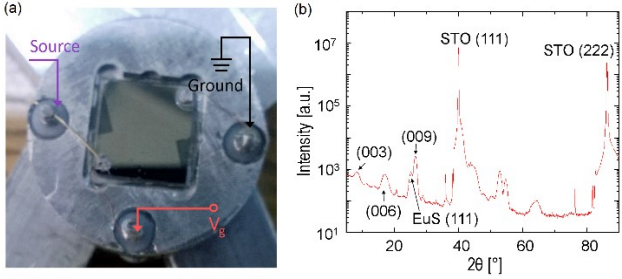

**Fig. S1.** (a) Real image of the MI/TI heterostructure on STO substrate for gate-dependent PNR measurements. (b) The X-ray diffraction results of the 4QL TI $(Bi_{0.2}Sb_{0.8})_2Te_3$ sample on STO (111) substrate. The strong (00$n$), $n$=3,6,9 thin film peaks indicate the TI's single crystalline structure.

## II. Interpretation of Spin Asymmetry (SA)

Here we argue that even prior to any fitting process, the SA obtained from raw data tuned by gate voltage is already a strong indication of the proximity induced magnetism change, using the polarized neutron reflectometry theory introduced in Chapter 3 of Ref. [1]. Defining the thin film growth direction as $z$ direction, neutron nuclear scattering length density (NSLD) and magnetic scattering length density (MSLD) as $\rho_n$ and $\rho_m$, respectively, then the total scattering length density for the two spin components as a function of z can be written as

$$\rho_\uparrow(z) = \rho_n(z) + \rho_m(z)$$
$$\rho_\downarrow(z) = \rho_n(z) - \rho_m(z) \quad (1)$$

The corresponding polarized reflectivity for the two neutron spin components can be written as

$$R_{\uparrow/\downarrow} = \frac{1}{Q^2}\left|\int e^{iQz}\rho_{\uparrow/\downarrow}(z)dz\right|^2 \quad (2)$$

where $Q$ is the momentum transfer along z-direction. Further defining the Fourier transform of NSLD and MSLD as

$$\chi_n \equiv \int e^{iQz}\rho_n(z)dz$$
$$\chi_m \equiv \int e^{iQz}\rho_m(z)dz \quad (3)$$

Then the SA can finally be written down as

$$SA \equiv \frac{R_\uparrow - R_\downarrow}{R_\uparrow + R_\downarrow} = \frac{2\chi_n\chi_m}{\chi_n^2 + \chi_m^2} \quad (4)$$

In other words, SA is directly linked with both NSLD $\rho_n$ and MSLD $\rho_m$. Since gate voltage tuning is in general not expected to change density or thickness hence NSLD [2], the change of SA can be reasonably understood as originated from Fourier-transformed MSLD $\chi_m$ change, i.e. a change of magnetic structure. Since $\rho_m = \rho_m(z)$ is a local property, and magnetism of EuS and proximity layer occur at different spatial localization, we could decompose $\chi_m$ as a sum of contribution from EuS and proximity layer, i.e.

$$\chi_m \equiv \int e^{iQz}\rho_m(z)dz \approx \int e^{iQz}\rho_{m,EuS}(z)dz + \int e^{iQz}\rho_{m,Proximity}(z)dz$$

Furthermore, since EuS is a large bandgap insulator whose magnetization is also not expected to change up voltage gating, we then infer that the change SA is solely coming from the change of proximity layer.

This serves as the basis why we could conjecture that the change of SA observed in experiment is at least mainly coming from proximity effect, even without any fitting to the data. Now we could argue why the SA increase (Fig. 2b inset in main text) indicates an increase of magnetic proximity effect instead of decrease.

Since the majority of the system is non-magnetic (from calculated NSLD of materials, or see Fig. 3 directly), we expect $\chi_n \gg \chi_m$, then from Eq. (4), in this region we see that SA increases with $\chi_m$. Since the increase of $\chi_m$ is speculated to arise from proximity, the observed SA increase is thus inferred as increase of proximity, at least in the $V_g > 0$ region.

For the $V_g < 0$ region, due to the warming and re-cooling to the sample to eliminate STO ferroelectric effect, the NSLD of the sample hence $\chi_n$ is changed accordingly, thereby we do not intend to compare with $V_g > 0$ to $V_g < 0$ (Fig. 2b inset).

### III. The scheme for high-quality fitting

To minimize the Figure of Merit (FOM) hence realize high-quality fitting and, we divide the refinement process into two sub-processes. For the coarse fitting (Fig. S2a), we start from a reasonable guess of the sample SLD, thickness and roughness, where the FOM is reduced. As long as the fitted thickness or SLD is relatively fixed, a fine fitting (Fig. S2b) incorporating the proximity magnetism is adopted to further reduce the FOM. A typical fitting curve is shown in Fig. S2c, where with this approach, the fitting error (normalized reflectivity difference between fitted value and measured value) can be suppressed at all Q range (Fig. S2c lower panel).

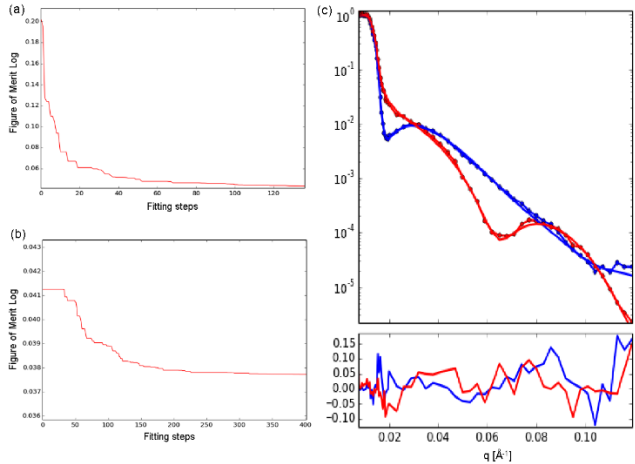

**Fig. S2.** The procedure of fitting. (a) Coarse optimization (b) Fine optimization and (c) Typical fitting curve (upper) and fitting error (lower).

To make different voltages more comparable, the FOM is maintained $< 5 \times 10^{-2}$, indicating the high-quality of fitting.

### IV. The fixed-SLD fitting

The price to pay for the procedure above is to slightly change the SLD and thickness. A separate global fitting with strictly fixed SLD and thickness, though having worse FOM, shows similar trend of proximity as a function of gate voltage, as shown in Fig. S3. It can also be seen that as the gate voltage $V_g$ is increased, the magnetic proximity effect is reduced (Fig. S3 inset). The SLD at +20V is almost overlapping +50V result, while the +200V result also overlaps with +100V result.

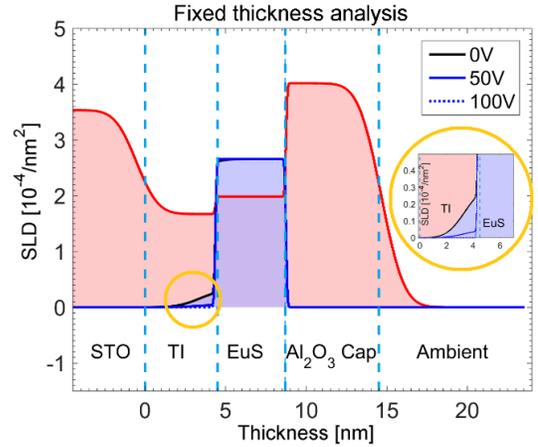

**Fig. S3.** The global fitting with fixed SLD and thickness across various gate voltages.

Due to the sample warming and re-cooling between $V_g > 0$ to $V_g < 0$, the SLD may experience a change together with the thickness, invalidating this global fitting scheme for $V_g < 0$ region.

[†]These authors contribute equally to this work.
[*] Authors to whom correspondence should be addressed:
mingda@mit.edu
gchen2@mit.edu
cxc955@psu.edu